\documentclass[11pt,a4wide]{article}
\usepackage{newlfont}
\usepackage{amsmath}
\usepackage{graphics}
\usepackage{float}
\usepackage{graphicx}
\usepackage{epsfig}

\begin{document}

\title{Higher Hybrid Bottomonia in an Extended Potential Model}
\author{Nosheen Akbar\thanks{%
e mail: nosheenakbar@ciitlahore.edu.pk} \quad,M. Atif Sultan \thanks{%
e mail: atifsultan.chep@pu.edu.pk} \quad, Bilal Masud\thanks{%
e mail: bilalmasud.chep@pu.edu.pk}\quad, Faisal Akram \thanks{%
e mail: faisal.chep@pu.edu.pk} \\
\textit{$\ast$COMSATS Institute of Information and Technology,
Lahore(54000), Pakistan.}\\
\textit{Centre For High Energy Physics, University of the Punjab,
Lahore(54590), Pakistan.}}
\date{}
\maketitle

\begin{abstract}
Using our extension of the quark potential model to hybrid mesons that fits well to the available lattice results,  we now calculate the masses, radii, wave functions at origin, leptonic and two photon decay widths, $E1$ and $M1$ radiative transitions for a significant number of bottomonium mesons. These mesons include both conventional and hybrid ones with radial and angular excitations. Our numerical solutions of the Schr$\ddot{o}$dinger equation are related to QCD through the Born-Oppenheimer approach. Relativistic corrections in masses and decay widths are also calculated by applying the leading order perturbation theory.  The calculated results are compared with available experimental data and the theoretical results by other groups. We also identify the states of $\Upsilon(10860)$, $\Upsilon(11020)$, and $Y_b(10890)$ mesons by comparing their experimental masses and decay widths with our results.
\end{abstract}


\section*{I. Introduction}

Models of Quantum chromodynamcis (QCD) can be tested by numerical (lattice) simulations of QCD or by hard experiments. We have earlier \cite{Nosheen11} suggested extensions to gluonic excitations of the analytical expressions for the quark-antiquark potential. We have good fits of the parameters in these expressions to the lattice simulations for the ground and \emph{excited state} gluonic field energy values available in ref.~\cite{Morningstar03} for discrete quark antiquark separations. In refs. \cite{Nosheen11,Nosheen14} we used these expressions to calculate a number of measurable quantities for
possible conventional and hybrid charomina. Now for the new sector of bottomonia we provide a comprehensive list of phenomenological implications that can be obtained through solving the Schr$\ddot{o}$dinger equation for our extended potential model; we report results for a variety of conventional and hybrid mesons. We include relativistic corrections and some other refinements to this non-relativistic treatment. For relating the angular momenta in our model with experimentally reported parity $P$ and charge parity $C$, we use the flux tube model \cite{Isgur1} that is applicable to hybrid quantum numbers as well. But the flux tube model was suggested only for large quark antiquark separations; we have earlier \cite{Nosheen11} pointed out that the additional potential term  ($\frac{\pi}{r}$) for hybrids resulting from this model differs significantly from the actual lattice simulations performed~\cite{morningstar}. Thus, instead of the flux tube model potential, we use our previously suggested quark potential model that has a very good comparison with these lattice-generated numbers. The relationship of quark potential model to QCD obtained through the adiabatic approximation and the Born-Oppenheimer formalism applies to our extended potential model as well; the already published~\cite{Nosheen11,Nosheen14,morningstar,Braaten} use of Born-Oppenheimer approach in hadronic physics includes gluonic excitations. Using a potential model, we are able to calculate masses, root mean square radii, and radial wave functions at origin of ground state and radially excited hybrid mesons as well. The radial wave function at origin is used in turn to calculate leptonic and two photon decay widths, and radiative transitions of these mesonic states. A comparison  of our results for bottomonium mesons of specified $J^{PC}$ states with experimentally known masses and decay widths with the same  $J^{PC}$ can help in identifying quarks and gluonic field configurations of the observed bottomonium mesons like $\Upsilon(10860)$, $\Upsilon(11020)$, and $Y_b(10890)$.

Heavy hybrid mesons have also been studied using theoretical approaches like constituent gluon model \cite{Iddir07,0611165,0611183}, QCD Sum Rule \cite{12040049,12094102,12066776,12083273,14106259,14110585,10122614,14037457}, lattice QCD  \cite{Braaten}, and Bethe-Salpeter equation \cite{9802360}.
 Wherever possible we compare with these theoretical results, along with available experimental numbers.

The paper is organized as follows. In the section 2, the potential models used for conventional and hybrid mesons are written. Then using these potential models, radial wave functions for the ground and excited states of conventional and hybrid bottomonium mesons are calculated by numerically solving the Schr$\ddot{o}$dinger equation. The relativistic corrections are subsequently included by the perturbation theory. The formulas used to calculate leptonic and two photon decay widths and radiative $E1$ and $M1$ transitions are also written in this section. Numerical results for these quantities for a variety of conventional and hybrid mesons are reported in section 3. Using our masses and leptonic decay widths for different $J^{PC}$ bottomonium states, we
also suggest possible assignments of $\Upsilon(10860)$, $\Upsilon(11020)$, and $Y_b(10890)$ mesonic states indicated through our extension of the quark model to hybrids.

\section*{2. Conventional and Hybrid Bottomonium mesons}

\bigskip

\subsection*{2.1. Spectrum of conventional mesons}

\noindent To study the spectrum of conventional bottomonium states we used the following semi-relativistic Hamiltonian including the lowest order relativistic correction%
\begin{equation}
H=2m_{b}+\frac{p^{2}}{2\mu }-\left( \frac{1}{4m_{b}^{3}}\right) p^{4}+V_{q%
\overline{q}}(r),
\end{equation}

\noindent where $\mu =m_{b}/2$ is the reduced mass of the system and $m_{b}$ is the constituent mass of the bottom quark. The effective quark anti-quark
potential $V_{q\bar{q}}(r)$ adopted from Ref.\cite{barnes05} carries Gaussian-smeared contact hyperfine interaction, one gluon exchange
spin-orbit and tensor terms, and the long ranged spin-orbit term in addition to the linear plus Coulombic terms. The complete expression of $V_{q\bar{q}}(r) $ is given by
\begin{eqnarray}
V_{q\bar{q}}(r) &=&\frac{-4\alpha _{s}}{3r}+br+\frac{32\pi \alpha _{s}}{%
9m_{b}^{2}}(\frac{\sigma }{\sqrt{\pi }})^{3}e^{-\sigma ^{2}r^{2}}%
\mathbf{S}_{b}.\mathbf{S}_{\bar{b}}  \notag \\
&&+\frac{1}{m_{b}^{2}}[(\frac{2\alpha _{s}}{r^{3}}-\frac{b}{2r})%
\mathbf{L}.\mathbf{S}+\frac{4\alpha _{s}}{r^{3}}T].
\end{eqnarray}

\noindent The Coulombic term which is proportional to strong coupling constant $\alpha _{s}$ arises from the one gluon exchange interaction dominating at short distance, whereas the linear term proportional to string tension $b$ is required to produce confinement in the system. The Gaussian-smeared contact hyperfine interaction proportional to $\mathbf{S}_{b}. \mathbf{S}_{\bar{b}}$, the short distance spin-orbit, and the tensor interactions are also produced by the one gluon exchange process, whereas the long ranged spin-orbit term is produced by the Lorentz scalar confinement. The tensor operator in the $\left\vert J,L,S\right\rangle $ basis is given by
\begin{eqnarray}
T& =\Bigg \{
\begin{array}{c}
-\frac{1}{6(2L+3)},J=L+1 \\
+\frac{1}{6},J=L \\
-\frac{L+1}{6(2L-1)},J=L-1.%
\end{array}%
\end{eqnarray}

\noindent Here $L$ and $S$ are quantum numbers of the relative orbital angular momentum of quark-antiquark and the total spin angular momentum of
the system respectively. Above effective potential carries four unknown parameters, strong coupling constant $\alpha _{s}$, string tension $b$, width
$\sigma $, and bottom quark constituent mass $m_{b}$. We fixed them by fitting the resulting spectrum to the experimental data composed of masses of ten well known
states of bottomonium mesons given in Table I. The best fitted values of these parameters, with relativistic correction, were found to be $\alpha _{s}= 0.4$, $b=0.11\text{ GeV}^{2}$, $\sigma =1$ GeV, $m_{b}=4.89$ GeV. Without the relativistic corrections, the best fitted values of parameters are $%
\alpha _{s}= 0.36 $, $b = 0.1340\text{ GeV}^{2}$, $\sigma = 1.34$ GeV, $%
m_{b} = 4.825$ GeV. To calculate the spectrum and the corresponding wave functions of the states of the  $b\overline{b}$ system we
numerically solved the radial Schr$\ddot{\textrm{o}}$dinger equation given by
\begin{equation}
U^{\prime \prime }(r)+2\mu (E-V(r)-\frac{\left\langle L_{q\bar{q}}^{2}\right\rangle}{2\mu r^{2}})U(r)=0, \label{de}
\end{equation}

\noindent where $U(r)=rR(r)$ with $R(r)$ being the radial part of the wave function and $\left\langle L_{q\bar{q}}^{2}\right\rangle = L(L+1)$. Non-trivial solutions of this equation, existing only for certain discrete values of $E$, were found by the shooting method. The above Schr$\ddot{\textrm{o}}$dinger equation assumes that the Hamiltonian $H=\frac{p^{2}}{2\mu}+V_{q\bar{q}}(r)$ i.e., without the constant term $2m_{b}$ and the relativistic correction. Thus to obtain the mass of a $b\overline{b}$ state we added the constituent quark masses to the above energy $E$ which was further corrected by the perturbation theory
for the lowest order relativistic correction to the Hamiltonian. That is,
\begin{equation}
m_{b\bar{b}}=2m_{b}+E+\left\langle \Psi \right\vert \Delta
H_{rel}\left\vert \Psi \right\rangle,
\end{equation}

\noindent where $\Delta H_{rel}=-\left( \frac{1}{4m_{b}^{3}}\right) p^{4}$
and $\Psi $ is the complete wave function of $b\bar{b}$ obtained by solving
the above Schr$\ddot{\textrm{o}}$dinger equation. It is noted that in limit $r\rightarrow 0$ the
potential $V_{q\bar{q}}(r)\sim \frac{2\alpha _{s}}{r^{3}}\left(
\mathbf{L}.\mathbf{S}+2T\right) $. It turns out that for $S=1$, $\mathbf{L}.\mathbf{S}+2T$ is
negative for $J=L$ and $J=L-1$. As a result, the potential becomes strongly
attractive at short distance and the resultant wave function becomes
unstable in this limit. To circumvent this problem, we calculated the meson
masses by solving Schr$\ddot{\text{o}}$dinger equation initially without
the spin-orbit coupling. The effect of spin-orbit interaction
was  subsequently included through the leading-order perturbative correction to the meson mass.
However, calculating the perturbative correction to the wave function is
difficult as in this case the contribution comes from all possible mass
eigen states. Therefore in this case we applied the smearing of position
coordinates, as discussed in Ref.~\cite{Isgur1}, to change the power behaviors of the potential at small distance.
 At small distance, smearing makes the potential less divergent than $1/r^{2}$. Thus
 with its use, the repulsive centrifugal potential
 $l(l+1)/(2\mu r^{2})$ remains dominating at small distance even if $\mathbf{L}.\mathbf{S}+2T$ is negative.

\subsection*{2.2.Characteristics of Hybrid Bottomonium mesons}

Our study of hybrid meson is based on Born-Oppenheimer (BO) approximation in which energy levels of gluonic field are
first calculated in the presence of static $q\overline{q}$ pair at fixed distance $r$ by Monte-Carlo estimates of generalized Wilson loops \cite{morningstar1}.
 These energy levels modify the effective $q\bar{q}$ potential $V_{q\bar{q}}(r)$ as following%
\begin{equation}
V_{q\bar{q}}^{h}(r)=V_{q\bar{q}}(r)+V_{g}(r),  \label{hypot}
\end{equation}

\noindent where $V_{g}(r)$ represents the contribution of the gluon field to
the effective potential. The functional form of $V_{g}(r)$ depends on the level of gluonic
excitation. In this work we study the hybrid mesons in which gluonic field
is in its first excited state and fit
the following form of $V_{g}(r)$
\begin{equation}
V_{g}(r)=\frac{c}{r}+A\times \exp^{-Br^{0.3723}}
\end{equation}
for a least difference of resulting gluonic excitation
 and ground state energies and the available lattice data~\cite{morningstar}.
We have shown in Ref.\cite{Nosheen11} that this proposed form of $V_{g}(r)$ provides an excellent fit to the lattice data
with best fitted values $A=3.4693$ GeV, $B=1.0110$ GeV, and $c=0.1745$. Thus, for the hybrid mesons, the form of the radial differential equation in Eq. (\ref{de}) is modified as
\begin{equation}
U^{\prime \prime }(r)+2\mu \left( E-V_{q\bar{q}}^{h}(r)-\frac{%
\left\langle L_{q\bar{q}}^{2}\right\rangle }{2\mu r^{2}}\right) U(r)=0, \label{eqU}
\end{equation}
\noindent where the squared quark anti-quark angular momentum $\left\langle L_{q%
\bar{q}}^{2}\right\rangle $ \cite{morningstar,Juge99} in leading Born-Oppenheimer approximation is given by
\begin{equation}
\left\langle L_{q\bar{q}}^{2}\right\rangle =L(L+1)-2\Lambda
^{2}+\left\langle J_{g}^{2}\right\rangle. \label{eq9}
\end{equation}

\noindent For the first gluonic excitation, the squared gluon angular momentum $%
\left\langle J_{g}^{2}\right\rangle =2$ and $\Lambda$, which is projection of gluon angular momentum $J_g$ on $q\bar{q}$ axis, is equal to 1 ~\cite{morningstar},
making $-2\Lambda ^{2}+\langle J_{g}^{2}\rangle =0$. In Eq. (\ref{eq9}) it is assumed that $\mathbf{L}=\mathbf{L}_{q\overline{q}}+\mathbf{J}_{g}$, which combines with total spin $\mathbf{S}$ of quark and antiquark to give total angular momentum $\mathbf{J}=\mathbf{L}+\mathbf{S}$ of a hybrid meson.

Using the hybrid potential of Eq. (\ref{hypot}), we calculated the masses and
radial wave functions of the hybrid mesons by using the same technique as employed for conventional mesons (mentioned above).
 The effect of the relativistic correction was again determined
using the leading-order perturbation theory. The resultant wave functions are
plotted in green color in Fig. 1 corresponding to same values of $n$,
$L$, and $S$ as used for conventional mesons.  The shape of these radial wave functions
is changed by the addition of the $V_{g}$ term for hybrids, and
the values of masses are significantly increased for the same values of $n$,
$L$, and $S$.
\begin{figure}
\begin{center}
\epsfig{file=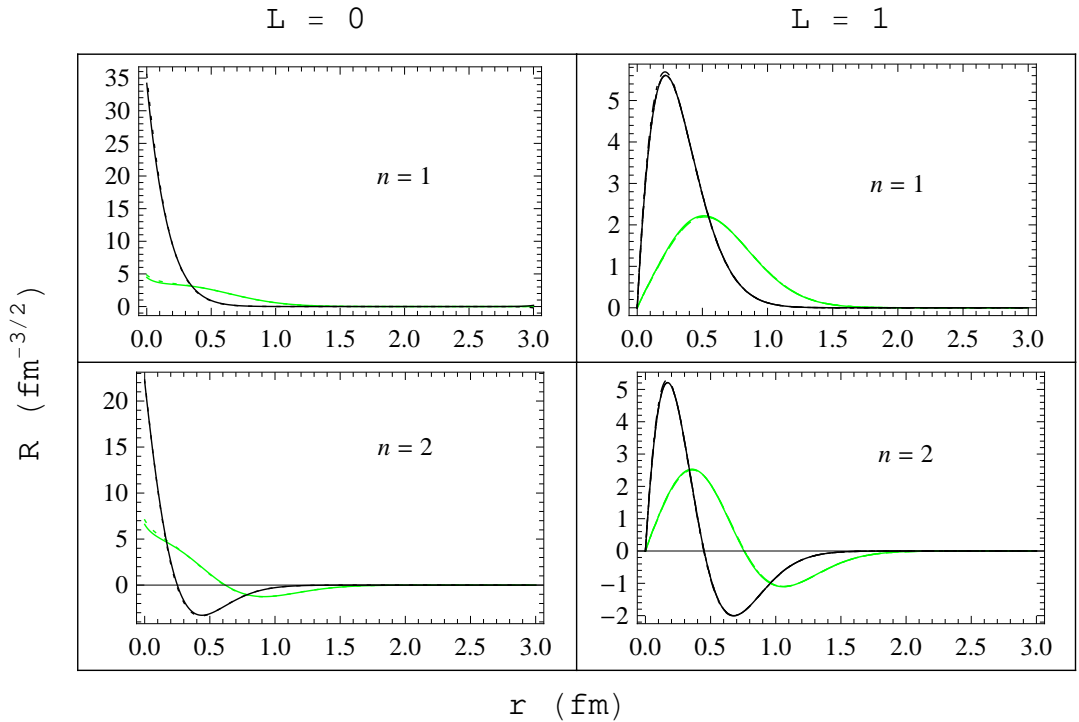,width=0.9\linewidth,clip=}\caption{Radial wave functions for ground and excited states. Black color represents conventional and green color represents hybrid bottomonium mesons. For $L=0$, wave functions of $\Upsilon$ and $\eta_b$ are almost same. Similarly for $L=1$, wave functions of $\chi_{b_0}$, $\chi_{b_1}$, $\chi_{b_2}$, and $h_b$ are almost same within our numerical limits and so on.}
\end{center}
\end{figure}
These normalized wave functions of conventional and hybrid bottomonium mesons were then used to calculate root mean square radii and radial wave functions at origin.
The applications of radial wave functions at origin are mentioned above in section 1.
The leptonic decay widths $\Gamma_{ee}$ of bottomonium mesons for $nS$ states were calculated by the following formula ~\cite{09091204}
\begin{equation}
\Gamma_{ee}(nS) = \frac{4 \alpha^2 e^2_b} {M^2_ns} \mid R_{nS}(0)\mid^2 (1 - \frac{16}{3} \frac{\alpha_s}{\pi}+\Delta(nS)). \label{l1}
\end{equation}
Here
 $e_b=-1/3$ is bottom quark electric charge and $\alpha$ the electromagnetic fine structure constant. The second term in parenthesis is the leading order radiative correction and $\Delta(nS)$
 stands for the higher order radiative and relativistic corrections. The value of $\Delta(nS)$ is state dependent. Following the Ref.~\cite{09091204}, we fixed it using the experimental value of $\Gamma_{ee}$ for $\Upsilon (4S)$, which is highest well known excited S-state. We found that for this state $\Delta(nS)=0.20$. However, the contribution of $\Delta(nS)$ is expected to be small for higher excited states as in this case relativistic effects are negligible \cite{09091204}.  So we take it zero for the excited states higher than $\Upsilon (4S)$.   For the $D$ states, leptonic decay widths were calculated by the formula \cite{09091369}
\begin{equation}
\Gamma_{ee}(n D) = \frac{25 \alpha^2 e^2_b} {2 M^2_nD M^4_b} \mid R''_{D}(0)\mid^2 (1 - \frac{16}{3} \frac{\alpha_s}{\pi}).
\end{equation}
WE calculated the two photon decay transitions for $S$ and $P$ states by using following expressions~\cite{09091369}
\begin{equation}
\Gamma(^1 S_0 \rightarrow \gamma \gamma) = \frac{3 \alpha^2 e^4_b \mid R_{nS}(0)\mid^2}{m^2_b},
\end{equation}
\begin{equation}
\Gamma(^3 P_0 \rightarrow \gamma \gamma) = \frac{27 \alpha^2 e^4_b \mid R'_{nP}(0)\mid^2}{m^4_b},
\end{equation}
\begin{equation}
\Gamma(^3 P_2 \rightarrow \gamma \gamma) = \frac{36 \alpha^2 e^4_b \mid R'_{nP}(0)\mid^2}{5 m^4_b}.  \label{2p3}
\end{equation}

Radiative transitions involve the emission of photon and are important for the study of mesons because they provide a way to access $b \overline{b}$ states with different internal quantum numbers $n$, $L$, $S$, $J$.  In $E1$ transitions orbital quantum numbers are changed but spin remain same. We calculated the $E1$ radiative partial widths from meson to meson transitions by using the following expression mentioned in Ref.~\cite{054034}.
\begin{equation}
\Gamma_{E1}(n^{2S+1}L_J\rightarrow n'^{2S+1}L'_{J'}+\gamma)=\frac{4 \alpha e_b^2 k_\gamma^3}{3}C_{fi}\delta_{L,L'\pm 1} \mid < \Psi_f \mid r \mid \Psi_i>\mid^2.  \label{E1}
\end{equation}
Here
 $k_\gamma$, $E^{b \overline{b}}_f$, and $M_i$ stand for final photon energy ($E_\gamma = \frac{M_i^2 - M_f^2}{2 M_i}$), energy of the final $b\bar{b}$ meson, and mass of initial state of bottomonium respectively, and
 \begin{equation}
C_{fi}=max(L, L')(2 J'+1)\left \{
                           \begin{array}{ccc}
                             J & 1 & J' \\
                             L' & S & L \\
                           \end{array}
                         \right \}.
\end{equation}
 To calculate the $M1$ radiative partial widths for meson to meson transitions, the following expression~\cite{054034} was used:
\begin{equation}
\Gamma_{M1}(n^{2S+1}L_J\rightarrow n'^{2S'+1}L_{J'}+\gamma)=\frac{4 \alpha e_b^2 k_\gamma^3}{3 m^2_b}\frac{2J'+1}{2 L+1}\delta_{S, S'\pm 1}\mid < \Psi_f \mid j_{0}(kr/2) \mid   \Psi_i>\mid^2. \label{M1}
\end{equation}
$j_{0}(x)$ is the spherical Bessel function.
We used Eqs. (\ref{E1}) and (\ref{M1}) to calculate both types of radiative transitions, i.e., conventional to conventional and hybrid to hybrid. However, E1 transition, which involves the change in $L$, requires some deliberation as $\mathbf{L}$ includes the contribution of gluon angular momentum $\mathbf{J}_g$. It is noted that the hybrids involved in E1 transition are calculated using the gluon potential
belonging to the first excitation of gluon field for which $\left\langle L_{q\bar{q}}^{2}\right\rangle=L(L+1)$ by Eq.(\ref{eq9}) meaning that the change in $L$ is caused by the corresponding change in $\left\langle L_{q\bar{q}}^{2}\right\rangle$.
 This gives, according to Eq.(\ref{eqU}), the spatial factors of wave functions of initial and final hybrid mesons appearing in Eq. (\ref{E1}). We do not need to change Eq. (\ref{E1}) because hybrid to hybrid transitions are between the same gluonic state, as are the conventional to conventional transitions; a flux tube model based analysis \cite{fluxtube}
of the $E1$ transitions involving hybrids replaces the quarks positions in the E1 transition operator by a separable combination of the relative position and a raising operator for the gluonic excitation.  This raising operator gives vanishing contribution to the transition amplitude between the hybrids belonging to same gluonic states and only the  old quarks relative position operator, producing transitions between conventional mesons, produces the E1 transitions between hybrid mesons.

\bigskip
\section*{3. Results and Discussion}
The masses, root mean square radii, radial wave functions at origin, leptonic and two photon decay widths, and radiative transitions are calculated for
conventional and hybrid bottomonium mesons including the orbital and radial excited $J^{PC}$ states. The calculated masses of bottomonium mesons using the non-relativistic and relativistic Hamiltonian for the ground and excited states are given in Table 1 along with the experimental and theoretical predictions of the other's groups. Table 1 shows that the lowest bottomonium meson's mass is $\approx 9.5$ GeV. Masses for hybrid bottomonium mesons using the non-relativistic and relativistic Hamiltonian for the ground state, orbital and radial excited states are reported in Table 2. For differentiating the symbols of conventional and hybrid mesons, a superscript $h$ is used with the corresponding conventional meson symbol having the same values of $n$, $L$, and $S$.
This is the same convention used earlier in our Ref.~\cite{Nosheen14}. In our calculations, the lowest hybrid bottomonium
has $J^{PC}=0^{++}(0^{--})$ with mass equal to $10.8069$ GeV
that is comparable to results mentioned in Ref.\cite{0611165}.
A comparison of Table 1 and 2 shows how much the mass of a hybrid meson is more than that of the corresponding conventional meson with same quantum numbers ($n$, $L$, and $S$). It is noted that $J^{PC}$ of each hybrid meson is different from the corresponding conventional meson for same $L$ and $S$. This difference is due to additional quantum numbers ($\Lambda $, $\varepsilon$, and $\eta$) present in the squared quark anti-quark angular momentum term for hybrid mesons defined in  Eq. (9) and in the following expressions for the parity ($P$) and charge parity ($C$) of
hybrid mesons.
\begin{equation}
P=\varepsilon (-1)^{L+\Lambda +1},C= \varepsilon \eta (-1)^{L+\Lambda +S},
\label{cp}
\end{equation}
\noindent where $\Lambda = 0, 1, 2, ....$ correspond to the states represented by symbols $\Sigma, \Pi_u, \Delta,...$ in Ref.~\cite{morningstar}. Here $\Sigma$
represents the mesons with ground state gluonic field, $\Pi_u$ corresponds to the mesons with first excited state gluonic field and so on. $\eta =-1$ and $\varepsilon =\pm 1$ for $\Pi _{u}$ state \cite{morningstar}. It is noted that parameter $\varepsilon$ has two possible values +1 and -1 for first gluonic excited state which we include in the study of hybrid mesons in our present work.  As a result we obtain two degenerate hybrid states with opposite values of $P$ and $C$. (This degeneracy has been recently discussed in Ref.~\cite{1510.04299} along with the nature of the approximation of using this in QCD.) For $\varepsilon =1$ the hybrid mesons are non-exotic, whereas exotic hybrid mesons are obtained for $\varepsilon =-1$
as shown in the Table 2.\\
The comparison of our calculated masses of bottomonium hybrid mesons with those by others is given in Table 3. We do not get as much differences in bottomonium hybrid masses predictions as lattice QCD and sum rules-based works report. But the constituent gluon model approach of Ref.~\cite{Iddir07} has differences comparable to ours. For $J^{PC}=1^{-+}$ we have an almost agreement with Ref.~\cite{Iddir07}. Our calculations give 10.8069 GeV for the mass of the lightest hybrid in the bottomonium sector without any radial excitation. This has $J^{PC}=0^{--}$. Ref.~\cite{Iddir07} reports the lightest hybrid meson to have a mass of 10.50 GeV with $J^{PC}=1^{--}$. The calculations reported in Refs. \cite{14037457} and \cite{13044522} using the QCD sum rule give lightest hybrid mass as low as 9.68 GeVwith $J^{PC}=0^{-+}$. This is part of their lightest hybrid supermultiplet. They take corresponding positive parity states belonging to the heavier hybrid supermultiplet starting from 10.17 GeV. But in the leading B.O. approximation, the $\Pi^+_u$ potential is same as that for $\Pi^-_u$ (\cite{morningstar, Braaten}) meaning that both parity states have the same mass. We use leading order B.O. approximation and thus have the same mass for both parities and both hybrid supermultiplets.  Ref.~\cite{Braaten} arbitrarily choose the mass of the $1^{--}$ member of their ground state $H_1$ multiplet, as defined in their Ref.46, to be the $B\bar{B}$ threshold i.e. 10.559 GeV and use the splittings from their Ref.14 to estimate the masses of the other bottomonium hybrids in lattice NRQCD.
These masses remain heavier than the chosen $1^{--}$ mass of 10.559 GeV. When they repeat the same procedure for the lattice QCD, this time choosing the $1^{-+}$ member of the ground state $H_1$ multiplet, they get the light bottomonium hybrid to be 10.159 GeV with $J^{PC}=0^{+-}$ having significant statistical errors of the lattice calculations and uncertainties from setting the heavy-quark mass.  Ref.\cite{Braaten} reports 440 MeV difference between the masses of the $1^{--}$ states in their radially ground and excited multiplets, which is significantly more than our corresponding difference of 128 MeV.
The quark confining string model has been used \cite{14097079} recently in QCD multipole expansion.  For the radially ground state (i.e. $n=1$), the only value they need and calculate,
their calculated mass  is 10.785 GeV with $J^{PC}=1^{--}$ which is not much different to our corresponding hybrid mass. A closely related vibrating string (flux tube) model used in \cite{REVD} has a nearly same value i.e. 10.789 GeV resulting from their model 3, but their model 1 gives a hybrid meson mass as 10.560 GeV.

Our root mean square radii and radial wave functions at origin for the ground and orbitally and radially excited states of conventional and hybrid bottomonium mesons are also reported in Tables 1 and 2 respectively.
We used the wave function at origin to calculate leptonic and two photon decay widths of conventional and hybrid bottomonium mesons using Eqs. \ref{l1} to \ref{2p3}.
Our calculated widths are given in Table 4 and 5 along with the corresponding available experimental data.
 The annihilation of the quarks takes place at the scale defined by the Compton wave length $1/ m_q$ which is close to zero for heavy quarks and hence the decay constant of a quarkonium state is proportional to square of its wave function at origin. Our results of wave functions at origin for conventional bottomonia given in Table 1 show that their values decrease with $n$, which implies that decay rates also decrease with increase in $n$ as shown in Table 4. This result is consistent with the experimental values of the decay rates given in the Table. However, in case of hybrid mesons the wave function at origin increases with $n$ as shown in Table 2. It is also noted that wave function at origin for a hybrid state is significantly lower than for the same conventional states. This makes decay widths of hybrids much lower than the conventional bottomonium mesons as shown in Table 4 and 5. Thus the hybrid states are characteristically different from conventional states as their leptonic and two photon decay rates are smaller and also increasing function of $n$ as shown in Table 4 and 5. This result can be very useful in finding the clues of hybrids in the experimental data.
Our results in Table 1 and 2 also show that for the same quantum numbers ($n$, $L$, and $S$) root mean square radii of hybrid mesons are greater than conventional mesons. It is also noted that radii of conventional and hybrid mesons increase with radial and angular excitations.

 Our calculated radiative magnetic dipole ($M1$) transitions and electric dipole ($E1$) transitions are reported in Tables (6-11) using the relativistic and non-relativistic masses. In the $M1$ transitions the initial and final states belong to the same orbital excitation but have different spins, and in the $E1$ transitions the orbital quantum numbers of initial and final states are changed but spins remain the same. Radiative transitions of excited bottomonium states are important because they provide a way to access $b\bar{b}$ states with different internal quantum numbers $(n,L,S,J)$. $\chi_b(1P)$ and $\chi_b(2P)$ states have been studied experimentally through the radiative decays of $\Upsilon(2S)$ and $\Upsilon(3S)$ states \cite{1112}. Recently ATLAS collaboration and LHCb reconstructed the $\chi_b(1P,2P,3P)$ states through the radiative decay $\chi_b(nP)\rightarrow\Upsilon(1S,2S)$ \cite{1112.5154}. It is noted that E1 Radiative transitions are typically of order of 1 to 10 keV, whereas M1 transitions are reduced due to the $m_{b}^2$ factor in the formula. Nevertheless M1 transitions have been useful in observing spin singlet states that are difficult to observe otherwise. It is also found that E1 transitions are suppressed for transition between the states that differ by 2 radial nodes like $5S\rightarrow3P$, $3P\rightarrow1D$, $4P\rightarrow2S$, $4D\rightarrow2S$, $4D\rightarrow2P$ and $3D\rightarrow1F$ \cite{1507}
 except the $3P\rightarrow1S$.
 We find same behavior in the case of radiative transitions of hybrid $b\bar{b}$ states. Generally both E1 and M1 transition rates are also very small when the transitions occur between the states with close masses because of reduced value of $E_{\gamma}$,  whereas the transition rates of strong decays are of order of few MeV \cite{054034}. As a result the branching  ratios of radiative transitions are significantly reduced when strong decays open at the threshold of $B\bar{B}$ states of energy $10.56$ GeV. Thus, branching ratios of radiative transition can be significantly high for the conventional 1S, 2S, 3S, 1P, 2P, 3P, 1D, 2D, 1F, and 1G states having mass less than 10.56 GeV. In case of hybrids states the branching ratios of radiative transitions are expected to be small because the mass of even lightest hybrid states is  greater than $B\bar{B}$ channel threshold  and hence competing strong decays are open for all hybrids. In Tables (6-11) the comparison of the decay widths of radiative transitions of conventional and hybrid $b\bar{b}$ states  with the available experimental data and results of refs.\cite{054034,d14} is provided. For most of the transitions the results are comparable to available experiments and other theoretical models. Reported values of decay widths of radiative transitions of hybrid $b\bar{b}$ states show that these values are smaller as compared to the values for corresponding transition for conventional states. These differences can be attributed to differences in the values of masses and radial wave functions.

Using our calculated masses and reported widths of conventional and hybrid mesons, we can identify the $\Upsilon(10860)$,$\Upsilon(11020)$, and $Y_b(10890)$ with help of the available experimental data.

\subsection*{(a). $\Upsilon(10860)$}

$\Upsilon(10860)$ has mass $10891 \pm 4$ MeV, $J^{PC}=1^{--}$, and leptonic decay width $0.31\pm 0.07$ keV \cite{pdg16}. The comparison of its experimental mass with calculated values of $1^{--}$ states shows that $\Upsilon(10860)$ could be $5S$ or $6S$ state of conventional bottomonium or $h^h_b(1P)$ hybrid state. However, Table 4 shows that calculated value of leptonic decay width of $5S$ state (0.23 keV) is closer to experimental value than that of $6S$ state. Since the leptonic decay widths of hybrid states are generally much lower than the conventional states, $\Upsilon(10860)$ is unlikely to be a hybrid states. Thus we suggest that $\Upsilon(10860)$ is $5S$ state of conventional bottomonium. Ref.\cite{14097079} also assigns $\Upsilon(10860)$ as $5S$ while ignoring the relativistic corrections, which are quite significant for bottomonia. In this reference the parameters of potential are fitted using $S$ states only, whereas we include 10 bottomonium mesons with different $S$, $P$, and $D$ states. On the other hand the Ref.\cite{09091204} assigns it $6S$ state using a potential which is screened at small distance by light $q\bar{q}$ pairs created in hadronic vacuum. However, their analysis is merely based on the comparison of mass values.

\subsection*{(b). $\Upsilon(11020)$}

$\Upsilon(11020)$ has mass $10987.5_{-3.4}^{+11}$ MeV, $J^{PC}=1^{--}$, and leptonic decay width $\Gamma_{ee}=0.130 \pm 0.030$ keV \cite{pdg16}. The comparison of its experimental mass and $J^{PC}$ with our results implies that it could be $6S$ state of conventional bottomonim or $2P$ hybrid state or $3P$ hybrid state. However, the possibility of this being hybrid is unlikely as the measured decay width is higher than that expected for the hybrids. As given in Table 4 the calculated value of $\Gamma_{ee}$ for $6S$ state is $0.20$ keV, assuming the relativistic correction term $\Delta(nS)=0.20$ (estimated from the experiment value of $\Gamma_{ee}$ of $4S$ state). As remarked earlier that the value of $\Delta(nS)$ is state dependent and expected to be small for higher excited states. If this correction term is ignored for $6S$ state then calculated $\Gamma_{ee}$ reduces to $0.128$ keV which is very close to experimental value.
Refs. \cite{09091369, 14097079} also assign $\Upsilon(11020)$ as $6S$ state of bottomonium meson contrary to Ref.\cite{09091204}. This reference assigns $\Upsilon(11020)$ to $7S$ state along with $\Upsilon(10860)$ to $6S$ state by considering $5S$ state as yet undiscovered.

\subsection*{(c). $Y_b(10890)$}
$Y_b(10890)$ has mass $10888.4 \pm 3$ MeV\cite{Belle}, $J^{PC}=1^{--}$, and leptonic decay width is yet not measured. The measured value of mass in this case closely corresponds to $5S$ state of conventional bottmonium and $1P$ hybrid state, both having $J^{PC}=1^{--}$. Since we have $5S$ state assigned to $\Upsilon(10860)$, therefore we suggest that $Y_b(10890)$ could be $1P$ hybrid state. Ref.\cite{1510.04299} also suggests that $Y_b(10890)$ is a candidate for bottomonium hybrid $1^{--}$ state and its decay to the $\Upsilon$ suggests that $Y_b(10890)$ is not a conventional bottomonium meson as this decay violates the spin symmetry. In Refs.~\cite{1012.2614,0812.3291}, estimated mass of $1^{--}$ bottomonium hybrid is compatible to the mass of $Y_b(10890)$. However, Refs. \cite{1011.2815} are in favour of $Y_b(10890)$ being a tetra-quark.

\section*{4. Summary}
We show that our lattice-based extension of the quark potential model to gluonic excitations allows us to conveniently calculate
the properties of both the conventional and hybrid bottomonium mesons without and with orbital and radial excitations.
 All this is
 needed to actively take part in the discussion for recognizing mesons in this sector. Relativistic corrections are included in our work. We find additional evidence to support the recent suggestion of assigning $Y_b(10890)$ to be $1P$ hybrid state. Our calculated conventional and hybrid masses and leptonic widths indicate that $\Upsilon(10860)$ and $\Upsilon (11020)$ are $5S$ and $6S$ states of bottomonium mesons.
We find how much hybrid meson masses and radii are more than those of the conventional mesons for the same $n,L$ and $S$ quantum numbers. Due to smaller values of the radial wave functions at origin for hybrids as compared to conventional meson states, our leptonic and photonic decay widths of hybrid states are much
smaller than the conventional states. It is also observed that in contrast to the trend for the conventional mesons, hybrid decay rates increase with principal quantum number. These can be useful signatures for recognizing hybrid mesons. Our values of $M1$ and $E1$ radiative transition for conventional bottomonium mesons have a
 good agreement with the corresponding experimental data. We find that for hybrid $b\bar{b}$ states the  values of the radiative decay widths are generally smaller than for conventional states. Given that the masses of hybrids states are always greater than $B\bar{B}$ threshold,
 we require high statistics to detect any signal for hybrids states notable through radiative transitions in present and future B factories.

Overall, this implementation indicates that our extended potential model can be used to advance predictions in a variety of meson sectors.

\begin{table}[tbp]
\caption{Masses, root mean square radii, and radial wave function at origin for ground, radial and orbital excited states of bottomonium mesons. Our
calculated masses are rounded to 0.0001 GeV.}
\begin{center}
\tabcolsep=1pt \fontsize{6.5}{6}\selectfont
\begin{tabular}{|c|c|c|c|c|c|c|c|c|}
\hline
Meson & \multicolumn{2}{|c|}{Our calculated mass} &  \multicolumn{2}{|c|}{Theoretical mass} & Experimental mass &Our calculated& Others Theor.& Our calculated  \\ \cline{2-3}\cline{4-5}
 & Relativistic & NR & Relativistic \cite{054034}& NR & & $\sqrt{ \langle r^{2} \rangle}$ & calculated $\sqrt{ \langle r^{2} \rangle}$& $|R(0)|^2$\\
 & & & potential model & potential model &~\cite{pdg14} & & ~\cite{09091369}&\\ \hline
 & \textrm{GeV} & \textrm{GeV} & \textrm{GeV} & \textrm{GeV} & \textrm{GeV} & fm & fm & $\textrm{GeV}^3$ \\ \hline
$\eta_{b} (1 ^1S_0)$ & 9.4926 & 9.5079 &9.402& 9.448 \cite{olka},9.428~\cite{holographic} &$9.399\pm 0.0023 $&0.2265 &- &11.8099 \\
 $\Upsilon (1 ^3S_1)$ & 9.5098 & 9.5299 & 9.465 & 9.459 \cite{olka},9.460~\cite{holographic} & $9.4603\pm 0.00026$ &0.2328 &0.23 &11.4185\\ \hline
$\eta_{b} (2 ^1S_0)$ & 10.0132 & 10.0041 & 9.976 & 10.006\cite{olka}, 10.190~\cite{holographic} & $9.999_{-1.9} ^{+2.8}$ &0.5408 &- &4.0509\\
$\Upsilon (2 ^3S_1)$ & 10.0169 & 10.0101 & 10.003& 10.009\cite{olka}, 10.219~\cite{holographic} & $10.02326 \pm 0.00031$ &0.5448 &0.52 &4.0393\\ \hline
$\eta_{b} (3 ^1S_0)$ & 10.2792 & 10.2912 & 10.336 & 10.352 \cite{olka},10.372~\cite{holographic} & - & 0.8018&- &2.9327 \\
$\Upsilon (3 ^3S_1)$ & 10.2815 & 10.295 & 10.354 & 10.354 \cite{olka},10.401~\cite{holographic} & $10.3552 \pm 0.0005$ & 0.8047 & 0.78 &2.9265 \\ \hline
$\eta_{b} (4 ^1S_0)$ & 10.4854 & 10.5214 & 10.623 & 10.473~\cite{holographic} & - &1.0273 & -&2.4758\\
$\Upsilon (4 ^3S_1)$ & 10.4872 & 10.5244 & 10.635 & 10.502~\cite{holographic} & $10.5794 \pm 0.0012$ & 1.0296&1.02 &2.4705 \\ \hline
$\eta_{b} (5 ^1S_0)$ & 10.6626 & 10.7226 & 10.869& --- & ---&- &- &2.2167 \\
$\Upsilon (5 ^3S_1)$& 10.6642 & 10.7251 & 10.878 & --- & &1.2324 &1.24 &2.2120 \\ \hline
$\eta_{b} (6 ^1S_0)$ & 10.8219 & 10.9053 & 11.097 &--- & ---& -&- &2.0452 \\
$\Upsilon(6 ^3S_1)$ & 10.8233 & 10.9074 & 11.102 &--- & --- & 1.4195&1.45 &2.0408 \\ \hline
$\eta_{b} (7 ^1S_0)$ & 10.9686 & 11.0748 & & --- & ---& & &1.9211 \\
$\Upsilon(7 ^3S_1)$ & 10.9698 & 11.0767 & &--- & --- & -& -&1.9170 \\ \hline
$h_{b} (1 ^1P_1) $ & 9.9672 & 9.9279 & 9.882 & &$ 9.8993 \pm 0.0001$ & 0.4347&- &0\\
$\chi_{0} (1 ^3P_0)$ & 9.8510 & 9.9232 & 9.847&9.871 \cite{olka},10.1160~\cite{holographic} & $9.85944\pm 0.00042 \pm 0.00031$ &0.4375 &- &0 \\
$\chi_{1} (1 ^3P_1)$ & 9.9612 & 9.9295 &9.876& 9.897 \cite{olka},10.190~\cite{holographic} & $9.89278 \pm 0.00026 \pm 0.00031 $& 0.4379&- &0\\
$\chi_{2} (1 ^3P_2)$ & 9.9826 & 9.9326 & 9.897& 9.916 \cite{olka},10.219~\cite{holographic} & $9.91221 \pm 0.00026 \pm 0.00031$ & 0.4375& 0.42&0 \\ \hline
$h_{b} (2 ^1P_1) $ & 10.2342 & 10.2213 & 10.250 & & ---& 0.7114& -&0 \\
$\chi_{0} (2 ^3P_0)$ & 10.2098 & 10.2197 & 10.226& 10.232 \cite{olka},10.343~\cite{holographic} & $10.2325 \pm 0.0004 \pm 0.0005$& 0.7132&- &0 \\
$\chi_{1} (2 ^3P_1)$ & 10.2306 & 10.2232 & 10.246& 10.255\cite{olka},10.372~\cite{holographic} & $10.25546 \pm 0.00022 \pm
0.00050$ & 0.7139&- &0\\
 $\chi_{2} (2 ^3P_2)$ & 10.2447 & 10.2245 & 10.261& 10.271\cite{olka},10.401~\cite{holographic} & $10.26865 \pm 0.00022 \pm
0.00050$ & 0.7139& 0.69 & 0\\ \hline
$h_{b} (3 ^1P_1) $ & 10.4423 & 10.456 & 10.541 & 10.444~\cite{holographic} &  & 0.9453 & -&0 \\
$\chi_{0} (3 ^3P_0)$ & 10.4239 & 10.4557 & 10.522 &10.522 \cite{olka},10.473~\cite{holographic} & --- &0.9470 & & 0\\
$\chi_{1} (3 ^3P_1)$ & 10.4396 & 10.4579 & 10.538& 10.544\cite{olka},10.502~\cite{holographic} & --- & 0.9476& &0 \\
$\chi_{2} (3 ^3P_2)$ & 10.4507 & 10.4585 &10.550& 10.559\cite{olka} & $10.534 \pm 0.009$ &0.9474 & 0.93 &0 \\ \hline
$h_{b} (4 ^1P_1) $ & 10.6213 & 10.6606 & & 10.521~\cite{holographic} & ---& 1.1541 & -&0 \\
$\chi_{0} (4 ^3P_0)$ & 10.606 & 10.6607 & 10.775& 10.550~\cite{holographic} & --- &1.1557 & -&0 \\
$\chi_{1} (4 ^3P_1)$ & 10.619 & 10.6624 & 10.788& 10.579~\cite{holographic} & --- &1.1561 & -&0 \\
$\chi_{2} (4 ^3P_2)$ & 10.6284 & 10.6627 & 10.798&--- & --- &1.1559 & -&0\\ \hline
$h_{b} (5 ^1P_1) $ & 10.7822 & 10.8459 & 10.790& --- & --- &1.3457 & -&0\\
$\chi_{0} (5 ^3P_0)$ & 10.7688 & 10.8463 &11.004& --- & ---& 1.3472& -&0 \\
$\chi_{1} (5 ^3P_1)$ & 10.7802 & 10.8476 &11.014 &--- & ---&1.3475 & -&0 \\
$\chi_{2} (5 ^3P_2)$ & 10.7884 & 10.8478 &11.022& --- & ---&1.3473 & 1.37&0 \\ \hline
$h_{b} (6 ^1P_1) $ & 10.9302 & 11.0176 & 11.016 &--- & --- &1.5246 & -&0\\
$\chi_{0} (6 ^3P_0)$ & 10.9182 & 11.0181 && --- & ---&1.5261 & -&0 \\
$\chi_{1} (6 ^3P_1)$ & 10.9284 & 11.0192 && --- & ---& 1.5263&-&0 \\
$\chi_{2} (6 ^3P_2)$ & 10.9358 & 11.0192 & & --- & ---&1.5260 & -&0 \\ \hline
$\eta_{b2} (1 ^1D_2)$ & 10.1661 & 10.1355 & 10.148& & ---&0.5933 & -&0 \\
$\Upsilon (1 ^3D_1)$ & 10.1548 & 10.1299 &10.138& & & 0.5930& -&0 \\
$\Upsilon_{2} (1 ^3D_2)$ & 10.1649 & 10.1351 & 10.147& 10.155~\cite{Braaten} & $10.1637 \pm 0.00014$ &0.5939 & -&0\\
$\Upsilon_{3} (1 ^3D_3)$ & 10.1772 & 10.1389 &10.155& & ---& 0.5942& -&0\\ \hline
\end{tabular}
\end{center}
\end{table}
\begin{table}[tbp]
\begin{center}
\tabcolsep=4pt \fontsize{9}{11}\selectfont
\begin{tabular}{|c|c|c|c|c|c|}
\hline
Meson & \multicolumn{2}{|c|}{Our calculated mass}& Relativistic potential &Our calculated& Others Theor.\\ \cline{2-3}
  & Relativistic & NR & model\cite{054034} &$\sqrt{ \langle r^{2} \rangle}$ & calculated $\sqrt{ \langle r^{2} \rangle}$\\
 &  & & & & \cite{09091369} \\ \hline
& \textrm{GeV} & \textrm{GeV}& \textrm{GeV}& fm & fm \\ \hline
$\eta_{b2}(2 ^1D_2)$ & 10.3801 & 10.3779 &10.450&0.8447 & -\\
$\Upsilon (2 ^3D_1)$ & 10.3688 & 10.3751&10.441&0.8448 &- \\
$\Upsilon_{2} (2 ^3D_2)$ & 10.3789 & 10.378& 10.449&0.8455 &-\\
$\Upsilon_{3} (2 ^3D_3)$ & 10.3864 & 10.3799& 10.455 &0.8457 & 0.82\\ \hline
$\eta_{b2}(3 ^1D_2)$ & 10.5633 & 10.5877& 10.706 & 1.0634 &-\\
$\Upsilon (3 ^3D_1)$ & 10.5525 & 10.5861&10.698 &1.0638 &- \\
$\Upsilon_{2} (3 ^3D_2)$ & 10.5621 & 10.5881& 10.705 &1.0643 &- \\
$\Upsilon_{3} (3 ^3D_3)$ & 10.5695 & 10.5892 & 10.711 &1.0643 & 1.05\\ \hline
$\eta_{b2}(4 ^1D_2)$ & 10.7273 & 10.777 & 10.935 & 1.2612&-\\
$\Upsilon (4 ^3D_1)$ & 10.717 & 10.776 & 10.928 & 1.2623&-\\
$\Upsilon_{2} (4 ^3D_2)$ & 10.7262 & 10.7775& 10.934 & 1.2626&- \\
$\Upsilon_{3} (4 ^3D_3)$ &10.7334 & 10.7782 & 10.939 &1.2625 &1.27\\ \hline
$\eta_{b2}(5 ^1D_2)$ &10.8779 & 10.9518 & &1.4455 &-\\
$\Upsilon (5 ^3D_1)$ & 10.8681 & 10.9512& & 1.4463 &-\\
$\Upsilon_{2} (5 ^3D_2)$ & 10.8768 & 10.9524& & 1.4465&- \\
$\Upsilon_{3} (5 ^3D_3)$ & 10.8839 & 10.9528 & &14463 &1.49 \\ \hline
$h_{b3}(1 ^1F_3) $ & 10.3116 & 10.2941&10.355 & 0.7280& -\\
$\chi_2 (1 ^3F_2)$ & 10.308 & 10.2894 &10.350 & 0.7266&-\\
$\chi_3(1 ^3F_3)$ & 10.3114 & 10.2937 &10.355 & 0.728&-\\
$\chi_4(1 ^3F_4)$ & 10.3137 & 10.297 &10.358 & 0.7289&-\\ \hline
$h_{b3}(2 ^1F_3) $ & 10.5 & 10.5102&10.619 & 0.9619&- \\
$\chi_2(2 ^3F_2)$ & 10.4956 & 10.5074 &10.615 & 0.9611 &-\\
$\chi_3(2 ^3F_3)$ & 10.4997 & 10.5101 &10.619 &0.9619 &-\\
$\chi_4(2 ^3F_4)$ & 10.5027 & 10.5119 &10.622 &0.9623 &-\\ \hline
$h_{b3}(3 ^1F_3) $ & 10.6679 & 10.7041&10.853 &1.1688 &- \\
$\chi_2(3 ^3F_2)$ & 10.6629 & 10.7022 &10.850 &1.1696 &-\\
$\chi_3(3 ^3F_3)$ & 10.6676 & 10.7041 &10.853 &1.1701 &- \\
$\chi_4(3 ^3F_4)$ & 10.6711 & 10.7053 &10.856 &1.1702 &- \\ \hline
$h_{b3}(4 ^1F_3) $ & 10.8216 & 10.8825& & 1.3621&-\\
$\chi_2(4^3F_2)$ & 10.8162 & 10.8811 & &1.3563 & -\\
$\chi_3(4 ^3F_3)$ & 10.8212 & 10.8826& &1.3567 &- \\
$\chi_4(4 ^3F_4)$ & 10.825 & 10.8835 & &1.3564 &-\\ \hline
\end{tabular}%
\end{center}
\end{table}

\begin{table}[tbp]
\caption{Our calculated masses, root mean square radii and radial wave function at origin for ground, radial and orbital excited states $b\overline{b}$ hybrid bottomonium mesons.}
\begin{center}
\tabcolsep=4pt \fontsize{9}{11}\selectfont
\begin{tabular}{|c|c|c|c|c|c|c|}
\hline
Meson &\multicolumn{2}{|c|}{$J^{PC}$} & \multicolumn{2}{|c|}{Our calculated mass}& our calculated&our calculated \\ \cline{2-3}\cline{4-5}
 &$\varepsilon=1$ & $\varepsilon=-1$ & Relativistic & NR & calculated $\sqrt{ \langle r^{2} \rangle}$& $|R(0)|^2$\\
\hline
& &  & \textrm{GeV} & \textrm{GeV} &fm &$GeV^3$\\ \hline
$\eta^{h}_{b} (1 ^1S_0)$ &$0^{++}$ & $0^{--}$ & 10.8069 & 10.7734 &0.6215&0.1445\\
$\Upsilon^{h} (1 ^3S_1)$ & $1^{+-}$ & $1^{-+}$ & 10.8079 &10.7747 & 0.6272& 0.1281 \\ \hline
$\eta^{h}_{b} (2 ^1S_0)$ & $0^{++}$ & $0^{--}$ & 10.9262 &10.9187 &0.874 &0.3448\\
$\Upsilon^{h} (2 ^3S_1)$ & $1^{+-}$ & $1^{-+}$ & 10.928 &10.9211 &0.8801 &0.3115 \\ \hline
$\eta^{h}_{b} (3 ^1S_0)$ & $0^{++}$ & $0^{--}$ & 11.0459 &11.0636 &1.0977&0.4873 \\
$\Upsilon^{h} (3 ^3S_1)$ & $1^{+-}$ & $1^{-+}$ & 11.048 &11.0664 &1.1027&0.4520\\ \hline
$\eta^{h}_{b} (4 ^1S_0)$ & $0^{++}$ & $0^{--}$ & 11.1642 &11.2057&1.3000&0.5727 \\
$\Upsilon^{h} (4 ^3S_1)$ &$1^{+-}$ & $1^{-+}$ & 11.1662 &11.2086 &1.3039 &0.5425\\ \hline
$\eta^{h}_{b} (5 ^1S_0)$ & $0^{++}$ & $0^{--}$ & 11.2798 &11.3442 &1.4864&0.6245 \\
$\Upsilon^{h} (5 ^3S_1)$ & $1^{+-}$ & $1^{-+}$ & 11.2817 &11.3469 &1.4895&0.6003\\ \hline
$\eta^{h}_{b} (6 ^1S_0)$ & $0^{++}$ & $0^{--}$ & 11.3924 &11.4789&1.6606&0.6579 \\
$\Upsilon^{h} (6 ^3S_1)$ & $1^{+-}$ & $1^{-+}$ & 11.394 &11.4814&1.6632 & 0.6385 \\ \hline
$h^{h}_{b} (1 ^1P_1) $ & $1^{--}$ & $1^{++}$ & 10.8561 &
10.8357&0.7601&0 \\
$\chi^{h}_{0} (1 ^3P_0)$ &$0^{-+}$ & $0^{+-}$ & 10.8534 &
10.8325 & 0.7564 & 0\\
$\chi^{h}_{1} (1 ^3P_1)$ & $1^{-+}$ & $1^{+-}$ & 10.8559 &
10.8354 & 0.7594 & 0\\
$\chi^{h}_{2} (1 ^3P_2)$ & $2^{-+}$ & $2^{+-}$ & 10.8569 &
10.8366 & 0.7623 & 0\\ \hline
$h^{h}_{b} (2 ^1P_1) $ &$1^{--}$ & $1^{++}$ & 10.984 &
10.9889 & 1.0014 & 0 \\
$\chi^{h}_{0} (2 ^3P_0)$ & $0^{-+}$ & $0^{+-}$ & 10.9792 &
10.9868 & 0.9996 & 0\\
$\chi^{h}_{1} (2 ^3P_1)$ & $1^{-+}$ & $1^{+-}$ & 10.9835 &
10.989 & 1.0015 & 0 \\
$\chi^{h}_{2} (2 ^3P_2)$ & $2^{-+}$ & $2^{+-}$ & 10.9856 &
10.9897 & 1.0035 & 0\\ \hline
$h^{h}_{b} (3 ^1P_1) $ & $1^{--}$ & $1^{++}$ & 11.1074 &
11.1363 & 1.2126 & 0\\
$\chi^{h}_{0} (3 ^3P_0)$ &$0^{-+}$ & $0^{+-}$ & 11.1012 &
11.135 & 1.2115 & 0 \\
$\chi^{h}_{1} (3 ^3P_1)$ &$1^{-+}$ & $1^{+-}$ & 11.1066 &
11.1367 & 1.2131 & 0\\
$\chi^{h}_{2} (3 ^3P_2)$ & $2^{-+}$ & $2^{+-}$ & 11.1097 &
11.1372 & 1.2149 & 0\\ \hline
$h^{h}_{b} (4 ^1P_1) $ & $1^{--}$ & $1^{++}$ & 11.2266 &
11.2786 & 1.4048 & 0\\
$\chi^{h}_{0} (4 ^3P_0)$ & $0^{-+}$ & $0^{+-}$ & 11.2195 &
11.2777 & 1.4041 & 0 \\
$\chi^{h}_{1} (4 ^3P_1)$ &$1^{-+}$ & $1^{+-}$ & 11.2257 &
11.2791 & 0.7564 & 0 \\
$\chi^{h}_{2} (4 ^3P_2)$ & $2^{-+}$ & $2^{+-}$ & 11.2293 &
11.2795 & 1.4071 & 0\\ \hline
$h^{h}_{b} (5 ^1P_1) $ & $1^{1--}$ & $1^{++}$ & 11.3418
& 11.4161 & 1.5834 & 0\\
$\chi^{h}_{0} (5 ^3P_0)$ & $0^{-+}$ & $0^{+-}$ & 11.3342 &
11.4155 & 1.5829 & 0\\
$\chi^{h}_{1} (5 ^3P_1)$ & $1^{-+}$ & $1^{+-}$ & 11.3408 &
11.4167 & 1.5841 & 0\\
$\chi^{h}_{2} (5 ^3P_2)$ & $2^{-+}$ & $2^{+-}$ & 11.3448 &
11.417 & 1.5858 & 0\\ \hline
$h^{h}_{b} (6 ^1P_1) $ & $1^{--}$ & $1^{++}$ & 11.4534 &
11.5493 & 1.7514 & 0\\
$\chi^{h}_{0} (6 ^3P_0)$ & $0^{-+}$ & $0^{+-}$ & 11.4456 &
11.549 & 1.7512 & 0\\
$\chi^{h}_{1} (6 ^3P_1)$ & $1^{-+}$ & $1^{+-}$ & 11.4524 &
11.5501 &1.7521 & 0 \\
$\chi^{h}_{2} (6 ^3P_2)$ & $2^{-+}$ & $2^{+-}$ & 11.4566 &
11.5503 & 1.7538 & 0 \\ \hline
$\eta_{b2} (1 ^1D_2)$ & $2^{++}$ & $2^{--}$ & 10.9125 &
10.9053 & 0.8771 & 0\\
$\Upsilon^{h} (1 ^3D_1)$ & $1^{+-}$ & $1^{-+}$ & 10.9119 & 10.9032& 0.8726 & 0 \\
$\Upsilon^{h}_{2} (1 ^3D_2)$ & $2^{+-}$ & $2^{-+}$ & 10.9126 &
10.9052 & 0.8761 & 0\\
$\Upsilon^{h}_{3} (1 ^3D_3)$ & $3^{+-}$ & $3^{-+}$ & 10.9127 &
10.9063 & 0.8799 & 0 \\ \hline
$\eta^{h}_{b2} (2 ^1D_2)$ & $2^{++}$ & $2^{--}$ & 11.0409
& 11.0583 & 1.1058 & 0\\
$\Upsilon^{h} (2 ^3D_1)$ & $1^{+-}$ & $1^{-+}$ & 11.0395 & 11.0566 & 1.1025 & 0
\\
$\Upsilon^{h}_{2} (2 ^3D_2)$ &$2^{+-}$ & $2^{-+}$ & 11.0408 &
11.0582 & 1.1051 & 0\\
$\Upsilon^{h}_{3} (2 ^3D_3)$ & $3^{+-}$ & $3^{-+}$ & 11.0415 &
11.0591& 1.1080 & 0 \\ \hline
\end{tabular}%
\end{center}
\end{table}
\begin{table}[tbp]
\begin{center}
\tabcolsep=4pt \fontsize{9}{11}\selectfont
\begin{tabular}{|c|c|c|c|c|c|}
\hline
 Meson & \multicolumn{2}{|c|}{$J^{PC}$} &
\multicolumn{2}{|c|}{Our calculated mass} & \\ \cline{2-3}\cline{4-5}
 & $\varepsilon=1$ & $\varepsilon=-1$ & Relativistic & NR& calculated $\sqrt{ \langle r^{2} \rangle}$ \\ \hline
& &  & \textrm{GeV} & \textrm{GeV} &fm \\ \hline
$\eta^{h}_{b2} (3 ^1D_2)$ & $2^{++}$ & $2^{--}$ & 11.1639
& 11.2047 & 1.3087\\
$\Upsilon^{h} (3 ^3D_1)$ & $1^{+-}$ & $1^{-+}$ & 11.1618 & 11.2034 &1.3059 \\
$\Upsilon^{h}_{2} (3 ^3D_2)$ & $2^{+-}$ & $2^{-+}$ & 11.1638 &
11.2048 &1.3081\\
$\Upsilon^{h}_{3} (3 ^3D_3)$ & $3^{+-}$ & $3^{-+}$ & 11.165 &
11.2054 &1.3108\\ \hline
$\eta^{h}_{b2} (4 ^1D_2)$ & $2^{++}$ & $2^{--}$ & 11.2823
& 11.3457 & 1.4947\\
$\Upsilon^{h} (4 ^3D_1)$ & $1^{+-}$ & $1^{-+}$ & 11.2796 & 11.3446 &1.4922
\\
$\Upsilon^{h}_{2} (4 ^3D_2)$ & $2^{+-}$ & $2^{-+}$ & 11.2821 &
11.3458 & 1.4943\\
$\Upsilon^{h}_{3} (4 ^3D_3)$ & $3^{+-}$ & $3^{-+}$ & 11.2837 &
11.3463 & 1.4969 \\ \hline
$\eta^{h}_{b2} (1 ^1D_2)$ & $2^{++}$ & $2^{--}$ & 11.3965
& 11.4818 &1.6685 \\
$\Upsilon^{h} (5 ^3D_1)$ &  $1^{+-}$ & $1^{-+}$ & 11.3933 & 11.4809&1.6661
\\
$\Upsilon^{h}_{2} (5 ^3D_2)$ &$2^{+-}$ & $2^{-+}$ & 11.3962 &
11.4819 &1.6682\\
$\Upsilon^{h}_{3} (5 ^3D_3)$ &$3^{+-}$ & $3^{-+}$ & 11.3983 &
11.4823&1.6707 \\ \hline
$h^{h}_{b3}(1 ^1 F_3) $ &$3^{--}$ & $3^{++}$ & 10.9727
& 10.9787&0.9856 \\
$\chi^{h}_{2}(1 ^3 F_2)$ & $2^{-+}$ & $2^{+-}$ & 10.9729 &
10.9774 &0.9806\\
$\chi^{h}_{3}(1 ^3 F_3)$ & $3^{-+}$ & $3^{+-}$ & 10.9729 &
10.9787 &0.9847\\
$\chi^{h}_{4}(1 ^3 F_4)$ & $4^{-+}$ & $4^{+-}$ & 10.9724 &
10.9793 &0.9891\\ \hline
$h^{h}_{b3}(2 ^1 F_3) $ & $3^{--}$ & $3^{++}$ & 11.0995
& 11.1293 &1.2029\\
$\chi^{h}_{2}(2 ^3 F_2)$ &$2^{-+}$ & $2^{+-}$ & 11.0993 &
11.1281 &1.1989\\
$\chi^{h}_{3}(2 ^3 F_3)$ &$3^{-+}$ & $3^{+-}$ & 11.0996 &
11.1293 &1.2023\\
$\chi^{h}_{4}(2 ^3 F_4)$ &$4^{-+}$ & $4^{+-}$ & 11.0995 &
11.13 &1.206\\ \hline
$h^{h}_{b3}(3 ^1 F_3) $ & $3^{--}$ & $3^{++}$ & 11.221
& 11.2736 &1.3983 \\
$\chi^{h}_{2}(3 ^3 F_2)$ &$2^{-+}$ & $2^{+-}$ & 11.2204 &
11.2725 &1.3947\\
$\chi^{h}_{3}(3 ^3 F_3)$ &$3^{-+}$ & $3^{+-}$ & 11.221 &
11.2737 &1.3977\\
$\chi^{h}_{4}(3 ^3 F_4)$ &$4^{-+}$ & $4^{+-}$ & 11.2212 &
11.2742&1.4012 \\ \hline
$h^{h}_{b3}(4 ^1 F_3) $ & $3^{--}$ & $3^{++}$ & 11.3378
& 11.4126 &1.5777\\
$\chi^{h}_{2}(4 ^3 F_2)$ &$2^{-+}$ & $2^{+-}$ & 11.3369 &
11.4116 &1.5759\\
$\chi^{h}_{3}(4 ^3 F_3)$ & $3^{-+}$ & $3^{+-}$ & 11.3378 &
11.4126 &1.5785 \\
$\chi^{h}_{4}(4 ^3 F_4)$ &$4^{-+}$ & $4^{+-}$ & 11.3383 &
11.4131 &1.5817\\ \hline
\end{tabular}%
\end{center}
\end{table}

\begin{table}[tbp]
\caption{Masses (in $\text{GeV}$) of hybrid bottomonium mesons calculated by
other
 models along with our calculated results. Our results
are reported for $J^{PC}$ states with lowest orbital angular momentum (L).}
\begin{center}
\tabcolsep=2pt \fontsize{9}{11}\selectfont
\begin{tabular}{|c|c|c|c|c|c|c|c|c|}
\hline
$J^{PC}$ & \multicolumn{2}{|c|}{Our results} & \multicolumn{2}{|c|}{NRLQCD}
& LQCD & QCD Sum Rule & Constituent  & Quark model  \\ \cline{2-3}\cline{4-5}
state & Ground state & First radial & Ground state & First radial &  &  &
Gluon Model & and QCS  \\
&  & excited state & \cite{Braaten} & \cite{Braaten} & \cite{Braaten}
& \cite{14037457}\cite{13044522} & \cite{Iddir07} &  \cite{14097079} \\
 \hline
$0^{--}$ & 10.8069 & 10.9262 & - &  & - & $11.48\pm0.75$ & 10.66 &  \\
$1^{--}$ & 10.8561 & 10.984 & 10.559 & 10.977+0.041 &  & $9.7\pm 0.12$ &
10.50 & 10.785\\
$0^{-+}$ & 10.8534 & 10.9792 &  &  &  & $9.68\pm0.29$ & 10.68 & \\
$1^{-+}$ & 10.8079 & 10.928 &  &  & 10.559 & $9.79\pm0.22$ & 10.80 & \\
$1^{+-}$ & 10.8079 & 10.928 &  &  &  & $10.70\pm 0.53$ &  & \\
$0^{+-}$ & 10.8534 & 10.9792 &  & - & $10.159\pm 0.362 $ & $10.17\pm0.22$ & & \\
$1^{++}$ & 10.8561 & 10.984 & $10.597\pm0.065$ & - & - & $11.09\pm0.60$ & -
& \\
$0^{++}$ & 10.8069 & 10.9262 & $10.892\pm0.036$ & - & - & $11.20\pm0.48$ & -
& \\
$2^{+-}$ & 10.8569 & 10.9856 & - & - & $11.323\pm0.257$ & - & - & \\
$2^{-+}$ & 10.8569 & 10.9856 & - & - &  & $9.93\pm 0.21$ & - &  \\
$2^{++}$ & 10.9125 & 11.0409 & - & - &  & $10.64\pm0.33$ & - & \\
\hline
\hline
\end{tabular}%
\end{center}
\end{table}

\begin{table}
\tabcolsep=4pt \fontsize{9}{11}\selectfont
\par
\begin{center}
\caption{Our calculated leptonic decay widths.}
\begin{tabular}{|c|c|c|c|c|c|}
\hline
State & \multicolumn{1}{|c|}{Conventional (keV)} &  Experimental &\multicolumn{1}{|c|}{Hybrid (keV)}\\ \hline
$1 ^3 S_1$ & 1.574  & $1.34\pm 0.018$ & 0.0135  \\
$2 ^3 S_1$ & 0.496  & $0.612\pm 0.011$ & 0.0322  \\
$3 ^3 S_1$ & 0.337  & $0.443\pm 0.008$ & 0.0457  \\
$4 ^3 S_1$ & 0.272  & $0.272\pm 0.029$ & 0.0537  \\
$5 ^3 S_1$ & 0.230  & $0.31 \pm 0.07$ & 0.0582  \\
$6 ^3 S_1$ & 0.209  & $0.13 \pm 0.030$ & 0.0607 \\
$1 ^3 D_1$ & 0.241  & - & 0.0016  \\
$2 ^3 D_1$ & 0.314  & - & 0.0033  \\
$3 ^3 D_1$ & 0.195  & - & 0.0294  \\
 \hline
\end{tabular}%
\end{center}
\end{table}

\begin{table}
\tabcolsep=4pt \fontsize{9}{11}\selectfont
\par
\begin{center}
\caption{Our calculated two-photon Decay widths.}
\begin{tabular}{|c|c|c|c|c|}
\hline
State & \multicolumn{1}{|c|}{Conventional (keV)} & \multicolumn{1}{|c|}{Hybrid (keV)}\\ \hline
$1 ^1 S_0$ & 0.813  & 0.0119  \\
$2 ^1 S_0$ & 0.271  & 0.0257  \\
$3 ^1 S_0$ & 0.195  & 0.0348  \\
$4 ^1 S_0$ & 0.165  & 0.0403  \\
$5 ^1 S_0$ & 0.148  & 0.0438  \\
$6 ^1 S_0$ & 0.136  & 0.0461  \\
$1 ^3 P_0$ & 0.354  & 0.00497  \\
$2 ^3 P_0$ & 0.346  & 0.00216  \\
$3 ^3 P_0$ & 0.284  & 0.00530  \\
$1 ^3 P_2$ & 0.109  & 0.000548  \\
$2 ^3 P_2$ & 0.105  & 0.003512  \\
$3 ^3 P_2$ & 0.139  & 0.00971  \\ \hline
\end{tabular}%
\end{center}
\end{table}

\begin{table}[tbp]
\caption{$S\rightarrow P$, E1 radiative transitions. Experimental results with $\dag$ sign are
calculated by considering the experimental measured branching ratio\cite{pdg12} and total decay width calculated in ref. \cite{054034}. we use the experimental masses if known. Otherwise,
theoretically calculated masses are used given in  Table 1 and 2.}\tabcolsep=4pt \fontsize{9}{11}%
\selectfont
\par
\begin{center}
\begin{tabular}{|c|c|c|c|c|c|c|c|c|}
\hline
Transition & Initial & Final & \multicolumn{2}{|c|}{Our calculated} & exp \cite{pdg12} & Others Theo.\cite{054034}  & \multicolumn{2}{|c|}{Our calculated} \\
& Meson & Meson & \multicolumn{2}{|c|}{$\Gamma_{E1}$} &  calculated $\Gamma_{E1}$ &calculated $\Gamma_{E1}$ &  \multicolumn{2}{|c|}{$\Gamma_{E1}$ for Hybrid}
\\ \cline{4-5}\cline{7-8}
&  &  & NR & Relativistic & & & NR & Relativistic \\ \hline
&  &  & keV & keV & keV & keV & keV & keV \\ \hline
$2S\rightarrow 1P$ & $\Upsilon (2 ^3S_1)$ & $\chi_2(1^3P_2)$ & 2.92446 &
3.1721 & $2.29\pm0.23$& 1.88& 1.7016 & 1.1754 \\
& & $\chi_1(1^3P_1)$ & 2.8324 &3.0722 & $2.21\pm 0.22$  & 1.63&
1.0648 & 0.7353 \\
& & $\chi_0(1^3P_0)$ &1.8530 & 2.0099 & $1.22\pm0.16$ & 0.91 & 0.3919 & 0.2713\\
& $\eta_b(2 ^1S_0)$ & $h_b(1^1P_1)$ & 4.5379& 6.1059 & & 2.48 & 2.9962 & 2.0837 \\ \hline
$3S\rightarrow 2P$ & $\Upsilon (3 ^3S_1)$ & $\chi_2(2^3P_2)$ & 3.5085 &
3.9429 & $2.66\pm0.41$ &2.3 & 2.8476 &1.8071 \\
& & $\chi_1(2^3P_1)$ & 3.2114 & 3.60897 & $2.56\pm 0.34$ &1.91
& 1.7555 & 1.9948\\
& & $\chi_0(2^3P_0)$ & 1.9819 & 2.2273 & $1.2\pm 0.16$ & 1.03&
0.6362 & 0.4837 \\
& $\eta_b(3 ^1S_0)$ & $h_b(2^1P_1)$ & 3.3529 & 1.0037 & -  & & 4.9051 &3.2650
\\ \hline
$3S\rightarrow 1P$ & $\Upsilon (3 ^3S_1)$ & $\chi_2(1^3P_2)$ & 0.478 &
0.8894 &$0.20 \pm0.03$ & 0.45 & 0.00082 & 0.0001752 \\
& & $\chi_1(1^3P_1)$ & 0.3247 & 0.6041 & $0.018 \pm 0.010$ & 0.05& 0.00050 & 0.0001779 \\
& & $\chi_0(1^3P_0)$ & 0.1323 & 0.2461 & $0.055 \pm 0.010$  & 0.01& 0.00017 & 0.00003697 \\
& $\eta_b(3 ^1S_0)$ & $h_b(1^1P_1)$ &0.7162& 1.0104 & - & &0.00075 & 0.003845 \\
\hline
$4S\rightarrow 3P$ & $\Upsilon (4 ^3S_1)$ & $\chi_2(3^3P_2)$ & 16.1868 &
22.1972 & & $0.82$ & 3.7232 &2.1898 \\
& & $\chi_1(3^3P_1)$ & 9.8560 &17.0252 & & 0.84 & 2.2809 & 1.5411 \\
& & $\chi_0(3^3P_0)$ & 3.4652 & 7.7808 & & 0.48 & 0.8152 & 0.6656 \\
& $\eta_b(4 ^1S_0)$ & $h_b(3^1P_1)$ & 4.6883 &1.5278 & &1.24 & 6.3651 &4.1018
\\ \hline
$4S\rightarrow 2P$ & $\Upsilon (4 ^3S_1)$ & $\chi_2(2^3P_2)$ & 0.2995 & 0.5918 & - & & 0.00052 &0.002269 \\
&  & $\chi_1(2^3P_1)$ & 0.2029 &0.401 & - & & 0.00031 & 0.002348 \\
&  & $\chi_0(2^3P_0)$ &0.0826 & 0.1632& - & & 0.00011 & 0.000503 \\
& $\eta_b(4 ^1S_0)$ & $h_b(2^1P_1)$ &0.5538 & 0.5644 &  & &
0.0041 & 0.0150\\ \hline
$4S\rightarrow 1P$ & $\Upsilon (4 ^3S_1)$ & $\chi_2(^3P_2)$ & 0.2212 &
0.3857 & - & & 0.0033 & 0.006019 \\
&  & $\chi_1(^3P_1)$ & 0.144 & 0.251 & - & & 0.0021 & 0.006076\\
&  & $\chi_0(^3P_0)$ & 0.0549 & 0.0957 & - & & 0.00069 & 0.001244 \\
& $\eta_b(4 ^1S_0)$ & $h_b(^1P_1)$ & 0.361& 0.4644 & - & & 0.0146 & 0.01693\\
\hline
\end{tabular}%
\end{center}
\end{table}
\begin{table}[tbp]
\caption{1P and 2P, E1 radiative transitions.}\tabcolsep=4pt \fontsize{9}{11}%
\selectfont
\par
\begin{center}
\begin{tabular}{|c|c|c|c|c|c|c|c|c|}
\hline
Transition & Initial & Final & \multicolumn{2}{|c|}{Our calculated} & exp.& others Theo \cite{054034} & \multicolumn{2}{|c|}{Our calculated} \\
& Meson & Meson & \multicolumn{2}{|c|}{$\Gamma_{E1}$} &  $\Gamma_{E1}$ & $\Gamma_{E1}$ & \multicolumn{2}{|c|}{$\Gamma_{E1}$ for Hybrid}
\\ \cline{4-5}\cline{7-8}
&  &  & NR & Relativistic &  & & NR & Relativistic \\ \hline
&  &  & keV & keV & keV & keV & keV & keV \\ \hline
$1P\rightarrow 1S$ & $\chi_2 (1 ^3P_2)$ & $\Upsilon(1^3S_1)$ & 37.2672 & 32.4094 &$34.38^{\dag}$ & 32.8 &  0.8501 & 0.5231 \\
& $\chi_1(1^3P_1)$ &  & 32.8195 & 14.0823 & $32.544^{\dag}$&29.5 &  0.8018 & 0.4918 \\
& $\chi_0(1^3P_0)$ &  & 26.0114 & 11.1611 & $--^{\dag}$&23.8& 0.6928 & 0.4192 \\
& $h_b(1 ^1P_1)$ & $\eta_b(1^1S_0)$ & 22.8589 &23.0341 &$35.77^{\dag}$ &35.7 & 0.8515 & 0.5217 \\
\hline
$2P\rightarrow 2S$ & $\chi_2 (2 ^3P_2)$ & $\acute{\Upsilon_2}(2^3S_1)$ & 18.9917
& 17.9986 &$24.645^{\dag}$ & 14.3 & 1.7320 & 1.2147 \\
& $\chi_1(2^3P_1)$ &  & 16.1418 & 15.2977  &$23.283^{\dag}$ &13.3 & 1.6798 & 1.0871 \\
& $\chi_0(2^3P_0)$ &  & 11.8767 & 11.2557 &$0.0001196^{\dag}$ &10.9 & 1.5225 & 0.8543 \\
& $h_b(2 ^1P_1)$ & $\acute{\eta_b}(2^1S_0)$ & 12.793 & 13.065 & $40.32^{\dag}$ &14.1 & 1.8030
& 1.1985 \\ \hline
$2P\rightarrow 1S$ & $\chi_2 (2 ^3P_2)$ & $\Upsilon(1^3S_1)$ & 14.3444 & 16.5078 &$16.275^{\dag}$ & 8.4 & 0.0828 & 0.0803 \\
& $\chi_1(2^3P_1)$ &  & 13.6951 & 7.7764& $10.764^{\dag}$&5.5 & 0.0820 & 0.0775 \\
& $\chi_0(2^3P_0)$ &  & 12.6092 & 7.1582&$0.0000234^{\dag}$ & 2.5 & 0.0796 & 0.072 \\
& $h_b(2 ^1P_1)$ & $\eta_b(1^1S_0)$ & 10.4703 & 12.9422 &$18.48^{\dag}$ & 13.0 & 0.0612 & 0.0641 \\
\hline
$2P\rightarrow 1D$ & $\chi_2 (2 ^3P_2)$ & $\Upsilon_3(1^3D_3)$ & 2.8853 &
2.1582& &1.5 & 1.6931 & 1.2894 \\
&  & $\Upsilon_2(1^3D_2)$ & 1.0543 & 0.5806 & & 0.3 & 0.3144 & 0.2312 \\
&  & $\Upsilon(1^3D_1)$ & 0.07872 & 0.0493 & &0.03 & 0.0225 & 0.0158\\
& $\chi_1 (2 ^3P_1)$ & $\Upsilon_2(1^3D_2)$ & 3.8710 & 1.9464 & & 1.2 & 1.5334 &
1.0596\\
&  & $\Upsilon(1^3D_1)$ & 1.4630 & 0.8546 & &0.5 & 0.5484 & 0.3637 \\
& $\chi_0(2^3P_0)$ & $\Upsilon(1^3D_1)$ & 3.2108 & 1.5811 & & 1.0 &2.0300 &
1.2093 \\
& $h_b(2 ^1P_1)$ & $\eta_{2b}(1^1D_2)$ & 1.9096 & 1.0775 & &1.7 & 2.0379 & 1.4554
\\ \hline
\end{tabular}%
\end{center}
\end{table}
\begin{table}[tbp]
\caption{3P E1 radiative transitions.}\tabcolsep=4pt \fontsize{9}{11}%
\selectfont
\par
\begin{center}
\begin{tabular}{|c|c|c|c|c|c|c|c|}
\hline
Transition & Initial & Final & \multicolumn{2}{|c|}{Our calculated} & Others
Theo. & \multicolumn{2}{|c|}{Our calculated} \\
& Meson & Meson & \multicolumn{2}{|c|}{$\Gamma_{E1}$} & calculated $%
\Gamma_{E1}$ & \multicolumn{2}{|c|}{$\Gamma_{E1}$ for Hybrid} \\
\cline{4-5}\cline{7-8}
&  &  & NR & Relativistic & & NR & Relativistic \\ \hline
&  &  & keV & keV & keV & keV & keV \\ \hline
$3P\rightarrow 3S$ & $\chi_2 (3 ^3P_2)$ & $\Upsilon(3^3S_1)$ & 2.7868& 2.1708 &8.2 \cite{d14} &
2.5898 & 1.9761 \\
& $\chi_1(3^3P_1)$ &  & 2.7389 & 1.5024 & 7.4 \cite{d14} & 2.5356 & 1.6942 \\
& $\chi_0(3^3P_0)$ &  & 2.568 & 0.8133 & 6.1 \cite{d14}& 2.3569 & 1.2692 \\
& $h_b(3 ^1P_1)$ & $\eta_b(3^1S_0)$ & 2.53206 & 10.5692 & 8.9 \cite{054034} & 2.71248 & 1.9078 \\ \hline
$3P\rightarrow 2S$ & $\chi_2 (3 ^3P_2)$ & $\acute{\Upsilon}(2^3S_1)$ & 3.8179 & 4.4072 & 3.8 \cite{d14} & 0.02522 & 0.02428 \\
& $\chi_1(3^3P_1)$ &  & 3.8027 & 4.0832 &2.5 \cite{d14}  & 0.02505 & 0.0231 \\
& $\chi_0(3^3P_0)$ &  & 3.7472 & 3.6518 &1.2 \cite{d14}  & 0.02447 & 0.02107 \\
& $h_b(3 ^1P_1)$ & $\eta_b(2^1S_0)$ & 4.3418 & 4.45835 & 8.2 \cite{054034} & 0.0085 & 0.01186 \\
\hline
$3P\rightarrow 1S$ & $\chi_2 (3 ^3P_2)$ & $\Upsilon(1^3S_1)$ & 6.5022 & 4.087 & 3.9 \cite{d14}&
0.02377 & 0.0182 \\
& $\chi_1(3^3P_1)$ &  & 6.4914 &  3.9609  & 2.1 \cite{d14}& 0.02367 & 0.01762 \\
& $\chi_0(3^3P_0)$ &  & 6.4517 & 3.7867 & 0.6 \cite{d14} & 0.02335 & 0.01670 \\
& $h_b(3 ^1P_1)$ & $\eta_b(1^1S_0)$ & 6.05386 & 7.4577 & 3.6 \cite{054034} & 0.014362 & 0.01225 \\
\hline
$3P\rightarrow 2D$ & $\chi_2 (3 ^3P_2)$ & $\Upsilon_3(2^3D_3)$ & 2.7611 & 1.7238 & 1.5 \cite{054034} &
2.9803 & 2.2670 \\
&  & $\Upsilon_2(2^3D_2)$ & 0.5294 & 0.4278 & 0.32 \cite{054034} & 0.5507 & 0.41735 \\
&  & $\Upsilon(2^3D_1)$ & 0.03922 &  0.0422  &  0.027 \cite{054034}& 0.03897 & 0.02942 \\
& $\chi_1 (3 ^3P_1)$ & $\Upsilon_2(2^3D_2)$ & 2.5888 & 1.2959 & 1.1 \cite{054034} & 2.7018 & 1.8188\\
&  & $\Upsilon(2^3D_1)$ & 0.959694 & 0.02735 & 0.4 \cite{d14} & 0.9565 & 0.6427 \\
& $\chi_0(3^3 P_0)$ & $\Upsilon(2^3D_1)$ & 3.5427 &1.2941& 0.9 \cite{d14}& 3.5887 & 2.0013\\
& $h_b(3 ^1P_1)$ & $\eta_{2b}(2^1D_2)$ & 3.2641 & 1.8727 & 1.6 \cite{054034} & 3.5555 & 2.5194 \\ \hline
$3P\rightarrow 1D$ & $\chi_2 (3 ^3P_2)$ & $\Upsilon_3(1^3D_3)$ & 0.003086 & 0.00995  & 0.046 \cite{054034} &
0.01358 & 0.009085 \\
&  & $\Upsilon_2(1^3D_2)$ & 0.0007225 & 0.002046 & - & 0.00246 & 0.001625\\
&  & $\Upsilon_(1^3D_1)$ & 0.00005047 & 0.0001492  & 0 \cite{d14} & 0.000168 & 0.0001094 \\
& $\chi_1 (3 ^3P_1)$ & $\Upsilon_2(1^3D_2)$ & 0.003593 &  0.009114  & 0.080 \cite{054034} & 0.01222 & 0.0077515
\\
&  & $\Upsilon(1^3D_1)$ & 0.001255 & 0.0001334 & $7 \times 10^{-3}$ \cite{d14} & 0.00418 & 0.002611\\
& $\chi_0(3^3 P_0)$ & $\Upsilon(1^3D_1)$ & 0.004922 & 0.01129 & 0.2 \cite{d14} & 0.01636 & 0.009612 \\
& $h_b(3 ^1P_1)$ & $\eta_{2b}(1^1D_2)$ & 0.011687 & 0.01829 & 0.081 \cite{054034} & 0.01439 & 0.009124 \\ \hline
\end{tabular}%
\end{center}
\end{table}

\begin{table}[tbp]
\caption{1D and 2D, E1 radiative transitions.}\tabcolsep=4pt \fontsize{9}{11}\selectfont
\par
\begin{center}
\begin{tabular}{|c|c|c|c|c|c|c|c|}
\hline
Transition & Initial & Final & \multicolumn{2}{|c|}{Our calculated} & Others
Theo. & \multicolumn{2}{|c|}{Our calculated} \\
& Meson & Meson & \multicolumn{2}{|c|}{$\Gamma_{E1}$} & calculated $%
\Gamma_{E1}$ & \multicolumn{2}{|c|}{$\Gamma_{E1}$ for Hybrid} \\
\cline{4-5}\cline{7-8}
&  &  & NR & Relativistic &\cite{054034}& NR & Relativistic \\ \hline
&  &  & keV & keV & keV & keV & keV \\ \hline
$1D\rightarrow 1P$ & $\Upsilon_3 (1 ^3D_3)$ & $\chi_2(1^3P_2)$ & 31.2721 &
37.4118 & 24.3 & 2.1424 & 1.3347\\
& $\Upsilon_2 (1 ^3D_2)$ & $\chi_2(1^3P_2)$ & 5.48044 &8.0209 & 5.6 & 0.5108 &
0.3319 \\
&  & $\chi_1(1^3P_1)$ & 21.025 & 29.9376& 19.2& 1.6137 & 1.0501 \\
& $\Upsilon(1^3D_1)$ & $\chi_2(1^3 P_2)$ & 0.568022 & 0.8016 & 0.56 & 0.05196 &0.0355 \\
&  & $\chi_1(1^3 P_1)$ & 10.9584 & 15.0785 & 9.7 & 0.8220 & 0.5621 \\
&  & $\chi_0(1^3 P_0)$ & 21.5026 & 28.5597 &16.5 & 1.2419 & o.8539 \\
& $h_{b2}(1^1 D_2)$ & $h_c(1^1P_1)$ & 25.8402 & 37.5565 & 24.9& 2.1301 & 1.3757 \\
\hline
$2D\rightarrow 2P$ & $\Upsilon_3 (2 ^3D_3)$ & $\chi_2(2^3P_2)$ & 5.066356 &
6.26455 & 16.4& 2.9765 & 1.8549 \\
& $\Upsilon_2 (2 ^3D_2)$ & $\chi_2(2^3P_2)$ & 1.20268 & 1.2878 & 3.8 & 0.7157 &
0.4466\\
&  & $\chi_1(2^3P_1)$ & 5.06132 & 5.4054 &12.7 & 2.2132 & 1.4978 \\
& $\Upsilon(2^3D_1)$ & $\chi_2(2^3 P_2)$ & 0.123364 & 0.1075 &0.4 & 0.07410 &0.04621\\
&  & $\chi_1(2^3 P_1)$ & 2.61871 & 2.3302 &6.5 & 1.1466 & 0.07770\\
&  & $\chi_0(2^3 P_0)$ & 5.87955 & 5.3734 & 10.6& 1.6822 & 1.2922 \\
& $\eta_{b2}(2^1 D_2)$ & $h_c(2^1P_1)$ & 13.7231 & 11.658 & 16.5& 2.9664 & 1.9484
\\ \hline
$2D\rightarrow 1P$ & $\Upsilon_3 (2 ^3D_3)$ & $\chi_2(1^3P_2)$ & 2.90251 & 3.9918
& 2.6 & 0.03892 &0.0460\\
& $\Upsilon_2 (2 ^3D_2)$ & $\chi_2(1^3P_2)$ & 0.717137 & 0.9530 & 0.4 & 0.00961 &
0.01137 \\
&  & $\chi_1(1^3P_1)$ & 2.42052 & 3.2158& 2.6& 0.02930 & 0.0347 \\
& $\Upsilon(2^3D_1)$ & $\chi_2(1^3 P_2)$ & 0.0782558 &0.09939&0.02 & 0.001046 &
0.001238 \\
&  & $\chi_1(1^3 P_1)$ & 1.32166 & 1.6815 &0.9 & 0.01594 & 0.01887 \\
&  & $\chi_0(1^3 P_0)$ & 2.13601 & 2.7245 & 2.9& 0.02208 & 0.02628\\
& $h_{b2}(2^1 D_2)$ & $h_c(1^1P_1)$ & 3.31247 & 4.3669&3 & 0.03686 & 0.04394\\
\hline
$2D\rightarrow 1F$ & $\Upsilon_3 (2 ^3D_3)$ & $\chi_4(1^3F_4)$ & 1.5356 & 1.1737 & 1.7 & 1.5441 & 1.1421 \\
 & & $\chi_3(1^3F_3)$ &  & 0.1113 &0.16 & 0.1365 & 0.09658 \\
 & & $\chi_2(1^3F_2)$ &  & 0.003629 & $5 \times 10^{-3}$& 0.00409 & 0.002759 \\
& $\Upsilon_2 (2 ^3D_2)$ & $\chi_3(1^3F_3)$ & 1.56236 &0.9104 &  1.5& 1.4778 & 0.9903\\
 &  & $\chi_2(1^3F_2)$ &  & 0.13177 &0.21 & 0.1939 & 0.1311 \\
 & $\Upsilon(2 ^3D_1)$ & $\chi_2(1^3F_2)$ &  & 0.7497 & 1.6 & 1.6439 & 1.1141 \\
& $\eta_{b2}(2^1 D_2)$ & $h_{c3}(1^1F_3)$ & 1.73254 &1.0752 &1.8 & 1.6696
&1.19693 \\ \hline
\end{tabular}%
\end{center}
\end{table}
\begin{table}[tbp]
\caption{1F and 2F, E1 radiative transitions.}\tabcolsep=4pt \fontsize{9}{11}%
\selectfont
\par
\begin{center}
\begin{tabular}{|c|c|c|c|c|c|c|c|}
\hline
Transition & Initial & Final & \multicolumn{2}{|c|}{Our calculated} & Others
Theo. & \multicolumn{2}{|c|}{Our calculated} \\
& Meson & Meson & \multicolumn{2}{|c|}{$\Gamma_{E1}$} & calculated $%
\Gamma_{E1}$ & \multicolumn{2}{|c|}{$\Gamma_{E1}$ for Hybrid} \\
\cline{4-5}\cline{7-8}
&  &  & NR & Relativistic &\cite{054034}& NR & Relativistic \\ \hline
&  &  & keV & keV & keV & keV & keV \\ \hline
$1F\rightarrow 1D$ & $\chi_4 (1 ^3F_4)$ & $\Upsilon_3(1^3D_3)$ & 9.5648 &
10.9949 &18 & 3.50637 & 2.2932 \\
& $\chi_3(1^3F_3)$ & $\Upsilon_3(1^3 D_3)$ & 0.986547 &1.1616 & 1.9& 0.3801 & 0.2612\\
&  & $\Upsilon_2(1^3D_2)$ & 14.2317 & 12.348 & 16.7& 3.1809 & 2.1002 \\
& $\chi_2(1 ^3F_2)$ & $\Upsilon_3(1^3D_3)$ & 0.0357111 & 0.0430 & 0.07& 0.01441 & 0.0104\\
&  & $\Upsilon_2(1^3D_2)$ & 2.2958 & 2.01672 &2.7 & 0.5278 & 0.3675 \\
&  & $\Upsilon(1^3D_1)$ & 13.676 & 13.004 &16.4 & 3.09214 & 2.0543 \\
& $h_{b3}(1^1F_3)$ & $\eta_{c2}(1^1D_2)$ & 15.9777 & 13.2407 & 18.8& 3.5635 & 2.3504 \\ \hline
$2F\rightarrow 2D$ & $\chi_4 (2 ^3F_4)$ & $\Upsilon_3(2^3D_3)$ & 14.0995 &
10.4506 & 19.6& 4.2665 & 2.7635 \\
& $\chi_3(2^3F_3)$ & $\Upsilon_3(2^3 D_3)$ & 1.50403 & 1.07438 & 2.1 & 0.46022 &
0.3086 \\
&  & $\Upsilon_2(2^3D_2)$ & 12.561 & 10.3987 & 17.9 & 3.8245 & 2.559 \\
& $\chi_2(2 ^3F_2)$ & $\Upsilon_3(2^3D_3)$ & 0.0565314 & 0.03851 & 0.08 & 0.01749 &
0.01215 \\
&  & $\Upsilon_2(2^3D_2)$ & 2.06744 & 1.6423 &3 & 0.6361 & 0.4410 \\
&  & $\Upsilon(2^3D_1)$ & 11.9235 & 11.3486 & 17.5 & 3.6751 & 2.5432\\
& $h_{b3}(2 ^1F_3)$ & $\eta_{c2}(2^1D_2)$& 14.1418 & 11.3816 &19.9 & 4.2826 &2.8476
\\ \hline
$2F\rightarrow 1D$ & $\chi_4 (2 ^3F_4)$ & $\Upsilon_3(1^3D_3)$ & 1.05143 & 1.1283
& 1.4 & 0.008949 & 0.01506 \\
& $\chi_3(2^3F_3)$ & $\Upsilon_3(1^3 D_3)$ & 0.115071 & 0.1220 & $2 \times 10^{-3}$ & 0.000985 &
0.001676\\
&  & $\Upsilon_2(1^3D_2)$ & 1.15993 & 1.1003 & & 0.007997 & 0.01343\\
& $\chi_2(2 ^3F_2)$ & $\Upsilon_3(1^3D_3)$ & 0.00449888 & 0.004701 & 0.1 & 0.00003878 & 0.00006671\\
&  & $\Upsilon_2(1^3D_2)$ & 0.198759 & 0.1858 & 1.4 & 0.001377 & 0.002339\\
&  & $\Upsilon(1^3D_1)$ & 1.11751 & 1.0837 & 1.6 & 0.007637 & 0.01277\\
& $h_{b3}(2 ^1F_3)$ & $\eta_{c2}(1^1D_2)$ & 1.32504 & 1.2458 & 1.6 & 0.008846 &0.01485
\\ \hline
\end{tabular}%
\end{center}
\end{table}

\section*{Acknowledgement}

\qquad B. M. and F. A. acknowledge the financial support of Punjab
University for the projects (Sr. 215 PU Project 2014-15 and Sr. 220 PU
Project 2014-15). N. A. is grateful to Higher education
Commission of Pakistan for their financial support (No:
PM-IPFP/HRD/HEC/2014/1703).

\begin{table}
\caption{ M1 radiative transitions calculated by taking the experimental masses from PDG \cite{pdg14}}%
\tabcolsep=4pt \fontsize{7}{9}\selectfont
\par
\begin{center}
\begin{tabular}{|c|c|c|c|c|c|c|c|c|}
\hline
Transition & Initial & Final & \multicolumn{2}{|c|}{Our calculated} &Exp & Others
Theo. & \multicolumn{2}{|c|}{Our calculated} \\
& Meson & Meson & \multicolumn{2}{|c|}{$\Gamma_{M1}$} & calculated $%
\Gamma_{M1}$ & calculated $%
\Gamma_{M1}$ &\multicolumn{2}{|c|}{$\Gamma_{M1}$ for Hybrid} \\
\cline{4-5}\cline{8-9}
&  &  & NR & Relativistic & &\cite{054034}& NR & Relativistic \\ \hline
&  &  & keV & keV & keV  & keV & keV & keV \\ \hline
$1S$ & $ \Upsilon (1 ^3S_1)$ & $\eta_b(1^1S_0)$ & 0.0110347 & 0.0002302 & & 0.010 & $1.01993\times 10^{-7}$ & $4.5204 \times
10^{-8}$ \\ \hline
2S & $\Upsilon(2^3 S_1)$ & $\acute{\eta}_b(2^1S_0)$ & 0.0006584 & 0.00004591 & & $5.9 \times 10^{-4}$& $ 6.4135\times 10^{-7}$ & $2.6351
\times 10^{-7}$  \\
&  & $\eta_b(1^1S_0)$ & 0.01729 & 0.0008832 &$0.012\pm 0.004$ & 0.081 &$0.000022269$ & $9.3836\times 10^{-6}$ \\
& $\eta_b(2 ^1S_0)$ & $\Upsilon (1 ^3S_1)$ & 0.0006363 & 0.003021 & & 0.068 & 0.0002045 &
0.00002694 \\ \hline
3S & $\Upsilon(3 ^3S_1)$ & $\eta_b(3^1S_0)$ & 0.011914 & 0.01948 & & $2.5\times 10^{-4}$& $1.01803\times 10^{-6}$ &$4.1833 \times 10^{-7}$ \\
&  & $\eta_b(2^1S_0)$ & 0.0030422 & 0.000151 &$<0.12$ &0.19 & 0.00004661 &  0.00001922 \\
&  & $\eta_b(1^1S_0)$ & 0.01718 & 0.0007183 & $0.01 \pm 0.002$&0.060 & 0.00004924 & 0.0000196 \\
& $\eta_b(3 ^1S_0)$ & $\acute{\Upsilon}(2^3S_1)$ & 0.0000107 & 0.0001993 &$0.949 \pm 0.098$ & 0.95 & 0.00013129 &  0.00005409\\
&  & $J/\Upsilon(1^3S_1)$ & 0.000116988 & 0.001816& $1.335 \pm 0.125$& 1.34& 0.0001390 & 0.00005519 \\
\hline
2 P & $h_{b}(2 ^1P_1)$ & $\chi_{2}(1^3P_2)$ & 0.0002621 & 0.0001772&& $2.2 \times 10^{-3}$ & $4.1391\times 10^{-7}$ &  $3.6028 \times 10^{-7}$ \\
&  & $\chi_{1}(1^3P_1)$ & 0.0002684 & 0.0001261 & & $1.1 \times 10^{-3}$ & $2.5419\times 10^{-7}$ &$2.2125 \times 10^{-7}$ \\
&  & $\chi_{0}(1^3P_0)$ & 0.000198333 & 0.00005515 & & $3.2 \times 10^{-4}$ & $8.9566\times10^{-8} $& $7.8111 \times 10^{-8}$ \\
& $\chi_{2}(2^3P_2)$ & $h_{b}(1^1P_1)$ & 0.001620 & 0.0001566& &$2.4 \times 10^{-4}$ &
$2.5471\times 10^{-7}$ &  $2.26827 \times 10^{-7}$ \\
& $\chi_{1}(2^3P_1)$ & $h_{b}(1^1P_1)$ & 0.001323 & 0.0001409& & $2.2 \times 10^{-3}$ &
$2.5129\times 10^{-7}$ &$2.1607 \times 10^{-7}$  \\
& $\chi_{0}(2^3P_0)$ & $h_{b}(1^1P_1)$ & 0.0009181 & 0.000116 & & $9.7 \times 10^{-3}$& $2.4075\times 10^{-7}$ &  $1.9511\times 10^{-7}$ \\ \hline
\end{tabular}%
\end{center}
\end{table}
\end{document}